\documentclass[preprint,12pt,longtitle,authoryear]{elsarticle}
\usepackage{epic}
\usepackage{latexsym}
\usepackage{amsmath}
\usepackage{epsfig}
\usepackage{amssymb}
\usepackage{enumerate}
\usepackage{color}

\usepackage[mathscr]{euscript}

\newtheorem{theorem}{Proposition}

\newtheorem{proposition}[theorem]{Proposition}

\newcommand{\cQ}{{\mathcal Q}}

\newcommand{\cR}{{\mathcal R}}
\newcommand{\cC}{{\mathcal C}}

\newcommand\ring[1]{\mathaccent23{#1}}
\usepackage{lipsum}            % Dummytext
\usepackage{xargs}             % Use >1 optional parameter in a new commands
\usepackage[pdftex,dvipsnames]{xcolor}  % Coloured text etc.

\begin{document}
\title{Autocatalytic networks in cognition and the origin of culture}
\author{Liane Gabora and Mike Steel}
%% You've put much more work into this so I think your name should go first. (Perhaps we will do a follow-up for a Psychology journal on which my name would go first.)

\begin{abstract} 
 It has been proposed that cultural evolution was made possible by a 
    cognitive transition brought about by onset of the capacity for 
    self-triggered recall and rehearsal.    Here we develop a novel idea that models  
    of collectively autocatalytic networks, developed for understanding the 
    origin and organization of life, may also help explain the origin of the 
    kind of cognitive structure that makes cultural evolution possible.  
   In this setting, mental representations (for example, memories, concepts, ideas) play the role of `molecules', and `reactions' involve the evoking of one representation by another through remindings and associations. In the `episodic mind', representations are so coarse-grained (encode too few properties) that such reactions must be `catalyzed' by external stimuli. As cranial capacity increased, representations became more fine-grained (encoded more features), which facilitated recursive catalysis and culminated in free-association and streams of thought. At this point, the mind could combine representations and adapt them to specific needs and situations, and thereby contribute to cultural evolution. In this paper, we propose and study a  simple and explicit cognitive model that gives rise naturally to autocatylatic networks, and thereby provides a possible mechanism for the transition from a pre-cultural episodic mind  to a mimetic mind.
\end{abstract}

\maketitle

\date{today} 

\bigskip

\noindent {\em Affiliations:}
[LG] Department of Psychology, University of British Colombia, Okanagan Campus, Kelowna BC, Canada;

\noindent [MS] Biomathematics Resarch Centre, University of Canterbury, Christchurch, New Zealand (corresponding author). 

\bigskip 
\noindent{\em Email:} liane.gabora@ubc.ca,  mike.steel@canterbury.ac.nz

\bigskip


%\newpage
\section{Introduction}

We are surrounded by evidence of a cultural evolution process that is cumulative (new ideas build on old ones), and open-ended (there is no {\it a priori} limit on the generation of novelty). By `culture' we refer to extrasomatic adaptations---including behavior and technology---that are socially rather than sexually transmitted. In order to contribute in a meaningful and reliable way to cultural evolution, one must be able to develop and refine ideas by thinking them through (i.e., engage in a stream of abstract thought). Since the capacity for a stream of thought is not specific to a particular domain of knowledge or cognitive process, the origins of this capacity are not straightforwardly traced to a particular area or even neural circuit of the brain. We could possess all the relevant neuroscientific data, as well as the relevant archaeological and anthropological data, yet still not understand how the human mind became able to evolve culture. Data alone are insufficient to explain this; what is needed here is a theory.

Once humans could engage in abstract thought, we could combine concepts, draw analogies, look at situations from different perspectives, modify plans according to unforeseen circumstances, and adapt ideas to new conditions, tastes, and desires. In other words, abstract thought enabled us to modify mental representations in light of one another, and thereby `reshape' our web of understandings as a whole. However, to engage in abstract thought requires that these representations be within reach of one another (i.e., they must already be somehow related to one another in our mind).   
Thus, in attempting to formally model the conditions for the emergence of cultural evolution, we are faced with the problem of explaining how a complex system composed of mutually dependent parts could come into existence. 
Which came first: the associative links between mental representations (i.e., the `tracks' that a `train of thought' runs on)? Or did trains of thought cement the connections from one rung (i.e., one mental representation) to the next? We have a `chicken and egg problem'.
In short, the answer to the question of how we arrived at the capacity for a stream of thought is related to the question of how we acquired an integrated web of understandings, and how this came about is not straightforward. 

Theories of how life began also face a `chicken and egg' problem as to how a self-replicating structure composed of mutually dependent parts could come into existence. The improbability that such a structure could come about spontaneously has led to widespread support for the hypothesis that the earliest forms of life were autocatalytic molecular networks that evolved in a relatively haphazard manner without an explicit self-assembly code, through a non-Darwinian process involving self-organization and horizontal (lateral) transfer of innovation protocols \citep{farmeretal1986, gab06, kau1, new00, seg00, vet06, wac97}. 
% This process has been referred to as communal exchange \citep{vet06}.
\cite{kau1} showed that when polymers interact, their diversity increases, and so does the probability that some subset of the total reaches a critical point where there is a catalytic pathway to every member, a state Kauffman referred to as \emph{autocatalytic closure}. 
Although the term `closure' has several different meanings in mathematics, and it is sometimes used to mean a condition that bounds or limits the set, 
 Kauffman (following in the footsteps of \cite{erdosrenyi1960}) used the term to express the property that the set surpassed a threshold density of connectedness by way of catalysis events. Thus, many closed sets within a system can exist, and they can become larger by combining together or through the introduction of new items.  Kauffman showed that autocatalytic sets emerge for a wide range of hypothetical chemistries (i.e., different collections of catalytic molecules). 
% next 2 sentences were moved up from next subsection

This paper explores the feasibility of adapting a generalized autocatalytic approach to model the emergence of the kind of cognitive structure necessary for complex culture. In other words, we draw upon a body of work developed to model the origin of life to model the transition to the kind of cognitive structure responsible for the origins of cultural evolution. While this paper is not the first to broaden the concept of `catalysis' and apply it in a cognitive context (see, for example, \cite{gab98, gab13, cat14}), here we build on these efforts to develop a more formal model that allows for analysis and predictions.

\subsection{Comparison of origins of biological and cultural evolution}
Although the origin of the kind of chemical structure necessary for biological evolution, and the origin of the kind of cognitive structure necessary for cultural evolution would appear superficially to be two very different problems, at a formal, algorithmic level they share a common deep structure. 
They both involves processes in which elements interact, generally referred to as \emph{reactions}, resulting in element transformation. In the case of the beginnings of biological evolution and the origin of life (OOL), the elements are catalytic molecules. In the case of the beginnings of cultural evolution and the origin of complex cognition (OOC), the elements participating in `reactions' are thoughts, memories, concepts, ideas, and chunks of knowledge encoded in memory which are referred to collectively as {\it mental representations}, or MRs. 
We use the term mental representation in what philosophers refer to as the `weak sense', in that we do not aim to provide a theory of consciousness.
In this paper, the term `reaction' will be used to refer to the process in which two or more MRs interact and at least one of them changes as a result. 
Although this paper does not delve into the mechanisms underlying this cognitive form of reactivity, we believe it arises due to overlap of receptive fields in distributed, content-addressable representations, as discussed in detail elsewhere (\cite{gab02, gab10, gabinpress, gabran13}).
The term `reactant' will be used to refer to the MRs participating in such a reaction. 

It is useful to distinguish between externally-driven and internally-driven reactions. In the OOC case, we use the term {\it learning} to refer to the cognitive process of revising a MR on the basis of new external input from the environment. We use the term {\it reflection} to refer to a cognitive process of revising a MR on the basis of internal input from the mind. 
% I hope you dont mind too much that i made this change of terminology. 
The mapping of the basic elements of OOL scenario to the OOC scenario is summarized in Table~1. 
In both the OOL and the OOC, certain elements, referred to as {\it catalysts}, speed up or help certain reactions along. In the OOL these elements are catalytic molecules, and in the OOC they are catalytic MRs. 
% (e.g., representational redescription, concept combination, etc.) 
% Thought it was better not to use these terms until they have been explained.
% Reorganized this part a bit so that reader has information needed to understand the table when it is encountered.

%\liane{I'm not sure i see why you're saying a CCP is an autocatalytic set. Perhaps that line should say "Closure / autocatalytic closure / conceptual closure? Isn't a CCP just a stream of thought? You can have a stream of thought without having achieved closure, isnt that right?}
\begin{table}[ht]
  \begin{center}
  \begin{tabular}{@{} clccccrrr @{}}
    \hline \hline 
    Component & OOL & OOC\\ 
    \hline
    % Moved this up 
    Food set & original polymers & innate concepts \\
     \hline
     Reactants & polymers & mental representations \\
    \hline 
     Products & polymers & mental representations \\
     \hline
   Reactions & ligation and cleavage & learning and reflection \\
   \hline
   Catalysts & polymers & stimuli and catalytic mental representations \\
   \hline
Autocatalytic sets & chemical RAFs & cognitive RAFs\\
\hline
  \end{tabular}
  \end{center}
\vspace{1mm}{\refstepcounter{table}\label{tab:variants}Table \arabic{table}. Comparison of the basic components of the two evolutionary processes that we propose are productively understood in terms of autocatalyic models. OOL refers to `origin of life' and OOC refers to `origin of culture'.} 
% Deleted next sentence because they havnt learned yet hat mimetic mind is.
% Updating can be catalyzed by a stimulus or (in the mimetic mind) by another mental item.}
%lg-feb9 changed 'reminding' to 'representational redescription' since to explain why reminding alters the representation would take another sentence or two.
\end{table}

Despite these similarities between the OOL and OOC scenarios, there are some important differences, which present interesting challenges. For example, in the OOC scenario, externally registered stimuli are held in working memory, whereas there is no similar bottleneck in the OOL. Such differences pose interesting theoretical challenges which are addressed in this paper.  
% For example, unlike the biochemical setting, here the sets are not necessarily subsets of some prescribed supersets (for example, the set of `all possible mental representations', as in \cite{gab01}). 
%\liane{This sentence might be confusing. Perhaps we could explain what the implications are of this distinction? Is it important to the model? If not can we leave it out? Also, i'm not completely sure i myself understand the relationship of the part in brackets to the first part f the sentence.}
%Second, in contrast to the diffuse and distributed nature of chemical networks, in the cognitive setting, catalyzed reactions are restricted by the `bottleneck' of attention, resulting in the content of working memory being continuously transformed (the much larger `inventory' of long-term memories provides a potential `food set' source for such catalysts). 

\subsection{Structure of paper}
% Building on earlier efforts \citep{gab98, gab01, gab13}, 
This paper sketches out the beginnings of a formal model of how the mind could have developed the kind of integrated structure that enables self-triggered recall and abstract thought, drawing upon a formal framework developed for the formal description of autocatalytic networks. 
The paper begins with a bare minimum of background material from psychology, anthropology, and archaeology concerning the uniqueness of human cognition and our ability to evolve complex, cumulatively creative culture.
Next, we provide the mathematical definitions of the basic concepts of our model, followed by the model itself. We then investigate the predictions of this model, in particular the factors that play an important role in the emergence of a kind of cognitive structure that is able to participate in cultural evolution, which we propose owes to the fact that is, in a fundamental sense, autocatalytic.
% from an episodic mind to a mimetic one, as well as the dynamics of autocatalytic networks, and collections of memories that attain `conceptual closure'. 
Finally, we conclude with some caveats and limitations, as well as unanswered questions and possibilities for future research. 

% I moved this next bit down because otherwise the layout of the paper didn't come until midway through the paper. 
\section{Background from cognitive anthropology: A transition in cognitive functioning}
We now summarize briefly the archaeological evidence that the origin of culture did reflect a transition to a different kind of cognitive functioning (see \cite{mit98,gab-prep,pen08,cho12, don91}). There is no consensus as to why {\it Homo erectus} crossed the threshold to the capacity for cumulative cultural evolution, but the cranial capacity of \emph{Homo erectus} was approximately 1,000 cc---about 25\% larger than that of \emph{Homo habilis} \citep{aie96}. Although simple stone (and some bone and antler) implements can be found in the archaeological record dating back to as long ago as 3.3 million years ago \citep{har15}, it is not until over a million years later that {\it Homo} constructed tools that were intentionally symmetrical \citep{lep11} and required multiple production steps and varied raw materials \citep{hai09}. By this time they were transporting tool stone over greater distances than their predecessors \citep{mou14}, and they had acquired the ability to use and control fire \citep{gor04}, had crossed stretches of open water up to 20 km \citep{gib98}, ranged as far north as latitude 52 \citep{par10}, revisited campsites possibly for seasons at a time, sometimes building shelters \citep{man05}, and ranked moderately high among predators \citep{plu04}. Thus, cumulative cultural evolution is believed to have originated approximately two million years ago, following the appearance of {\it Homo erectus} \citep{mit98}. 

It has been proposed that the increase in brain size enabled a transition to a fundamentally different kind of cognitive architecture \citep{don91}.\footnote{For related proposals see \citep{gab98, mit98, pen08}.}
\cite{don91} refers to the cognition of \emph{Homo habilis} as an \emph{episodic mode} of cognitive functioning because it was limited to the `here and now' of the present moment. He proposed that the enlarged cranial capacity enabled the hominin mind to undergo a transition to a new mode of cognitive functioning made possible by the onset of what he calls a \emph{self-triggered recall and rehearsal loop}, which we abbreviate STR. STR enabled hominins to voluntarily retrieve stored memories independent of environmental cues (sometimes referred to as `autocuing') and engage in pantomime, re-enactive play, and, importantly, representational redescription, which involves embellishing and revising thoughts and ideas as they are reflected upon from different perspectives. Donald referred to this as the \emph{mimetic mind} because it could act out or `mime' events that occurred in the past or that could occur in the future, thereby not only temporarily escaping the present but, through mime or gesture, communicating the escape to others. STR also enabled attention to be directed away from the external world toward ones' internal representations, which paved the way for abstract thought. It enabled systematic evaluation and improvement of thoughts and motor acts by adapting them to new situations, resulting in voluntary rehearsal and refinement of skills and artifacts. 

Donald's concept of STR bears some resemblance to the suggestion by \cite{hau02} that what distinguishes human cognition from that of other species is the capacity for recursion \citep{cor11}, as well as the concepts of relational reinterpretation by \cite{pen08} and of 
 `merge' by \cite{cho12} (for an overview, see \citep{gab-press}). 
What these theories have in common is that they focus not on abilities in a particular domain (such as social or technical abilities) but on a cognitive trait that cuts across domains. STR enabled not just the redescription and thereby refinement of previous representations but the sequential chaining of them, and in a way that, through autocuing, could build upon past representations. However, STR requires that concepts and ideas be accessible to one another (i.e., that they collectively constitute an integrated structure). How did such a structure emerge?

\section{Earlier approaches}
\label{early}
We now briefly review the earlier work upon which this paper builds. Inspired by Stuart Kauffman's  \citep{kau2, kau1} models of the emergence of the earliest kind of living structure (sometimes called a protocell) through autocatalytic closure of a set of catalytic polymers, \cite{gab98, gab00, gab01} proposed that the cognitive analog of the protocell is an individual's integrated web of knowledge, beliefs, and so forth, that constitute an internal model of the world, or worldview. 
% INTRODUCE M AND P 00L
In Kauffman's OOL model, each polymer,   composed of up to a maximum of $M$ monomers,  is assigned a low {\em a priori} probability $P$ of catalyzing each ligation (joining) and cleavage (cutting) reaction. 
% INTRODUCE M AND P 00C
In the cognitive scenario, the analog of the set of polymers is the set of MRs (i.e., mental representations in working or long-term memory).  The cognitive analog of $M$ (the maximum polymer length) is the maximum number of properties of a MR, and the analog of $P$, the probability of catalysis, is the probability that one representation brings about a reminding or associative recall of another.

%CLOSURE OOL AND OOC
It was proposed that, as exposure to similar items or events causes the formation of abstract concepts that connect these instances (for example, it is recognized that experiences of specific rocks are instances of the concept ROCK), a critical `percolation threshold' is eventually reached because the number of ways of forging associations amongst items in memory increases  faster than the number of items in memory. Following Kauffman's use of the term `autocatalytic closure' in early biochemistry, the analogous  state in cognition was referred to as \emph{conceptual closure} (Gabora, 2000), a term we will use later in this paper, with a precise definition. In this way, the assemblage of human worldviews changes over time not because some of them replicate in their entirety at the expense of others (for example, by  natural selection) but through piecemeal mutual interaction and self-organized transformation. Artifacts, rituals, and other elements of culture reflect the states of the worldviews that generate them. Interactions amongst items in memory increases their joint complexity, eventually transforming them into a conceptual network, which continually revises itself as new inputs are incorporated. This enables the creative connecting and refining of concepts and ideas necessary for the individual to participate in the evolution of cultural novelty.

% TRADEOFF M AND P FOR CLOSURE 00L
Kauffman found that, the lower the value of $P$, the greater $M$ has to be (and vice versa) in order for autocatalytic closure to be achieved. 
In other words, there is a transition from a subcritical process to a critical process which depends sensitively on these parameters. 
Similarly, there is a trade-off between (the analogues of) these parameters in conceptual closure. In the cognitive scenario, if the probability of associative recall is low, a network is subcritical: the worldview is expected to be stable but to have difficulty incorporating new information. Conversely, if the probability of associative recall is high, the network is super-critical: the worldview is expected to incorporate new information readily, but be at risk of destabilization (i.e., everything seems reminiscent of everything).

In the next few sections, we take these ideas in a new direction, and provide for a more explicit mathematical framework.  We stress once again that in the terms autocatalytic closure and conceptual closure, the word `closure' is not a condition that requires the set of items to be bounded or unable to grow larger. Rather it expresses a property that that the set is sufficiently connected together by catalysis events. In this way, many closed sets within a system can exist and they can become larger by combining together or by the introduction of new items.

\section{Background from Theoretical Work on Autocatalytic Sets}
\label{rafback}

As noted, the role of autocatalytic networks in OOL was introduced by \cite{kau1, kau2} in a pioneering approach to explain how complex biochemistry might have arisen from primitive chemistry, based on reactions that combine polymers.  The notion of self-sustaining autocatalytic sets  was developed further  (mathematically) as RAF-theory (here, RAF= Reflexively-autocatalytic and F-generated set, reviewed in \cite{hor16}).   

Formally,  a {\it catalytic reaction system} (CRS) is a tuple $\cQ=(X,\cR, C, F)$ consisting of a set $X$ of molecule types, a set $\cR$ of reactions, a catalysis set $C$ indicating which molecule types catalyze which reactions, and a subset $F$ of $X$ called the food set. 
A {\it Reflexively Autocatalytic and F-generated} (RAF) set for $\cQ$ is a non-empty subset $\cR' \subseteq \cR$ of reactions  which is:
\begin{enumerate}
  \item {\it Reflexively Autocatalytic}: each reaction $r \in \cR'$ is catalyzed by at least one molecule type that is either a product of $\cR'$ or is present in the food set $F$; and
  \item {\it F-generated}: all reactants in $\cR'$ can be created from the food set $F$ using a series of reactions only from $\cR'$ itself.
\end{enumerate}
In words, a RAF set is a subset of reactions that is both {\it self-sustaining} (i.e., every molecule involved in a reaction can be generated from the food set $F$ by a sequence of reactions within the subset) and collectively {\it autocatalytic} (i.e., every reaction is catalyzed by a molecule generated by the subset or the food set).  Such a set is a basic requirement for all living systems. \cite{kau1} first demonstrated this in a simple binary polymer model, the emergence of such a RAF occurs when the complexity of the polymers reaches a certain threshold. This has been further formalized and analyzed (mathematically and using simulations) and with applications to real biochemical systems (\cite{horetal10, horetal11}, \cite{horsteel04, horsteel12, hor16}, \cite{mos}). RAF theory has proven useful for identifying how such transitions might occur, and at what parameter values. 

Our approach here is to apply the theory of RAFs in a form that is maximally abstract, and show how this can be used to address the question of how a mind that is more or less a brittle, un-creative stimulus--response machine could transform into a mind capable of viewing situations from different perspectives, combining information from seemingly unrelated sources to solve problems, and adapting responses in contextually appropriate ways. We will start with this generic form of the model and examine how this might occur. We will see that as we incorporate aspects unique to the OOC scenario, the situation becomes more complex but the RAF approach can still effectively model it.

In short, although results concerning the emergence of RAFs in chemical networks cannot be applied directly to the question of how human cognition evolved, related mathematical techniques can be, as we show after introducing some further background material and definitions.

\section{A simple cognitive model based on reactions and catalysis}
\noindent At a top level, our highly simplified cognitive model can be viewed as a continuous-time, stochastic process involving three sets.
As mentioned in the introduction, we use the term {\em mental representation}, abbreviated MR, to refer collectively to items in memory (either working memory or long-term memory), as well as percepts, concepts, thoughts, and ideas, as well as more complex mental structures such as schemas. For simplicity, we view a MR to be an ensemble of a collection of hierarchically organized {\em properties}. 
% I tried to make this consistently organized, with the term at the beginning of the bullet point so that people can easily look back and see what a given term denotes.
\begin{itemize}
\item
$S_t$ denotes the set of stimuli at time $t$ registered by the senses (i.e., percepts that arise from sensory experiences). We can take $t=0$ to be the time of conception of the individual. 
%% I deleted the sentence below because (1) insofar as a subset of these items enters working memory they become internal to the mind, and (2) it could alienate those who view the senses are part of the mind. But if you'd rather leave it in let me know.
% Since $S_t$ is external to the mind, we make no attempt to model $S_t$.
\item
$L_t$ denotes the set of items encoded in long-term memory.  This includes the set
$I$ of any {\em innate knowledge} with which the individual comes into the world.
\item
$M_t$ denotes the set of items encoded in working memory and/or long-term memory at time $t$ (and so  $I \subseteq M_0$).\footnote{Not to be confused with $M$ from Section~\ref{early}.}
Each element of $m$ in $M_t$ is associated with a set of  properties, denoted $P(m)$, and we let $|m|$ be the number of those properties. Items in long-term memory encode largely static `invariances' of the world, while the items in working memory often reflect variances from this static model.

\item
$\ring{w_t}$ denotes the mental representation of a particular instant of experience at time $t$.  We will call $\ring{w_t}$ the {\em attended item} at time $t$. It is whatever is in the `spotlight' of attention at time $t$.
\item 
$W_t$ denotes items in {\it working memory.} It consists of $\ring{w_t}$ as well as any other similar or recently-attended items that are still accessible. It is a very small subset of $M_t$, of size in the order of $1$ to $\sim 10^3$. Thus, $\ring{w_t} \in W_t \subset M_t$.
% (here `recently' refers to  time in the order of minutes to hours, and so
\end{itemize}

There is a straightforward  way to define what `associated' means here (in the definition of $M_t$),  based on a natural  partial order on the set of  mental items.  For two mental items $m$ and $m'$ let us write $m \preceq m'$ if the properties comprising $m$ are a subset of the properties comprising $m'$. This partial order allows us to capture the notion of an item $m$ generalizing more particular instances $m_1, m_2, \ldots, m_k$ if $m\preceq m_i$ for each $i$ (for example, $m$ has precisely the properties shared by each of $m_1, \ldots, m_k$). 
The items $m$ and $m'$ are said be {\it members of a concept} if (though, not if and only if), on the basis of one or more shared properties, there exists a representation of an abstract category or prototype of which both are {\it instances}. For example, if $m=$ a smooth STONE, and $m'=$ a rough STONE, then $m'' =$ STONE is a lower bound (under $\preceq)$ to both of them.\footnote{Note that there are other ways that mental representations can be associated with each other (beyond sharing properties), such as via classical conditioning (for example, if a bell always goes right before food appears).}

More generally, the  partial order also allows for associations amongst items. We will say that $m$ and $m'$ are {\it associable} if there is some property they share, i.e., there exists $m''$ with $m'' \preceq m$ and $m'' \preceq m'$. Note that the properties associated with a mental representation are not fixed but can be biased by context or by mode of thought (i.e., convergent/analytic versus divergent/associative) (\cite{vel, sow, gabinpress}).
The fact that they share a particular property need never have been explicitly noted by the individual. For example, even if an individual has never consciously noticed that wood and rock share the property `hard', they are nevertheless implicitly associable. 
We will say that $m$ and $m'$ are {\it associated} if the fact that they share this property has been explicitly perceived and encoded in memory. For example, if an individual noticed that wood and rock share the property `hard', they would be  associated. 

The reason we stress the distinction between associable and associated is that explicit recognition of previously unrecognized shared properties is central to the creative processes that fuel cultural evolution. 
For example, one might consciously recognize that wood and rock share the property `hard'. If one had previously carved something durable out of wood, this would be a first step toward recognizing that a durable object might also be carved out of stone.

\subsection{Forging of cognitive structure}
\label{forging}
There are several sources of cognitive structure. One is innate structure in the form of instincts, fixed action patterns, and so forth. A second is structure on the basis of aspects of the MR perceived at the time of initial encoding, such as on the basis of properties shared by the MR and other previously encoded MRs. A third is structure on the basis of aspects of the MR perceived {\it after} the time of encoding, such as occurs during mind wandering, contemplation, reflection from different perspectives, or creative thought.
% revised last sentence very slightly - lg

As mentioned, the term {\it catalysis} refers to one MR evoking another, as in a stream of thought. A MR that was present at time $t$ plays the role of (and is referred to as) the {\it reactant}, whereas a MR that is present at time $t+\delta$ is referred to as the {\it product}. Catalysis may be precipitated by an external stimulus---as when something in the environment triggers a particular thought---or by another MR---as when the shift from one thought to another is triggered by looking at it from (one's internal representation of) the perspective of someone else. A stimulus or MR that precipitates cognitive catalysis is referred to as a {\it catalyst}.
We write the `reaction' $x \rightarrow z$ or $x\xrightarrow{y}z$, where $x$ is a {\em reactant} and $y$ is a {\em catalyst}.

In biochemistry, the distinction between reactant and catalyst is that the reactant is transformed (and therefore replaced by its product) in the reaction. The catalyst simply enables this to occur, without being itself used up in the reaction. In cognition, however, both the reactant and the catalyst may be affected by the reaction. For example, if $x$ is the MR of a piece of wood, and $y$ is the MR of an event in which the wood is dented by a falling rock, this may change the individual's conception of both wood (i.e., it is now dented) and rock (i.e., it is capable of denting wood). In the cognitive scenario, the distinction between reactant and catalyst is the following: the reactant is the MR that is generally (though not always) the focus of attention, whereas the catalyst is a stimulus or another MR that allows the reactant to give rise to a new MR as the next focus of attention.

\subsection{Modeling the episodic mind}
We posit that the dynamics of an episodic mind can be modeled by the following three processes. Note that in writing $+\delta$, we allow either a continuous-time process or a finely-discretized (i.e., nearly continuous time) process. 
\begin{itemize}

\bigskip

\item[]{\bf Encoding in memory}:
An item in $W_t$ can be encoded to long-term memory $m$ in $L_{t+\delta}$. We denote encoding by writing 
$w \leadsto m$. 
%\liane{Would it be simpler to assume for now that anything that is encoded in memory has been the focus of attention? It is actually more correct what you do here, i.e., there is evidence of subliminal encoding. But perhaps that is too ambitious for right now? I'm ok either way, but I'm leaning towards just mentioning subliminal processing as something for future work. That is why i revised this bit, but I'm happy to go back to how you had it if you think we should.}
and for an attended item $\ring{w}$, we write $\ring{w} \leadsto m$. Note that $w$ may or may not remain in $W_{t+\delta}$ when encoded to long-term memory. 

\bigskip

\item[]{\bf Shift in attention}:
% Made a few very minor changes here for readability - lg
Attention may shift from one thought or stimulus $\ring{w}$ in $W_t$ to another $\ring{w'}$ in $W_{t+\delta}$. 
After attention shifts away from a particular item $\ring{w}$, it persists for some time in $W_t$, and during this time it is still readily accessible. Once it is no longer present in working memory, it is denoted $w \mapsto \emptyset$.  Note, however, that it meanwhile may have been encoded in long-term memory.
$\ring{w'}$ may either come from working memory (in which case, we denote this shift in attention by writing $\ring{w} \mapsto w \mbox{ and } w' \mapsto \ring{w'}$)
or it may be a new item generated by one of the following processes.

\bigskip

\item[] {\bf Updating by stimuli}:
Without entering a detailed discussion on the nature of awareness and perception, (in order to keep the focus on the forest rather than the trees), we use the term {\it updating by stimulus} to refer to a stimulus-driven change in what is paid attention to, 
whether it be social in nature (such as a smile, gesture, or speech), or nonsocial (such as a change in the weather). Note that we are not using the term `learning' for this purpose because learning could imply that the change is encoded to long-term memory; what we are referring to here is any attended stimulus change, no matter how trivial, whether or not it is ever consolidated to long-term memory.

\indent We say that the subject of attention, $\ring{w}$, shifts to $\ring{w'} \in W_{t+\delta}$, due to catalysis by a stimulus $s$ in $S_t$; in other words: 
$$\ring{w} \xrightarrow{s} \ring{w'}  \mbox{ and } \ring{w} \mapsto w. $$
A concrete example of updating by stimuli is given in the lower part of Fig.~\ref{mimfig}.

\indent Sometimes the new content of working memory is not an external stimulus $s \in S_t$, but a mental representation 
$m \in M_t$ that was evoked by the stimulus because they are associated. This association may either have been hardwired or learned. For now, we will not concern ourselves with exactly how the stimulus affects the content of working memory, or the role of long-term memory (as well as goals, attitudes, motives, and so forth) on this process; what we focus on is the fact that the content of working memory has changed. Note that, in a society of interacting individuals, the expression, through speech or action, of an item $m \in M_t$ in one individual's mind can be regarded as a stimulus $s$ for another individual, thereby provide a social learning `reaction' in that individual. 
% Thus the collection of minds (together with other stimuli from the environment) provides a higher-level network structure.
% I think we should leave this comment out for now because it implies we are addressing that social level structure, but we are not, or at least not in this paper. Perhaps we could put it in the Discussion as a possibility for further work, mentioning the related projects of modeling economic networks.
\end{itemize}
% The above processes take place in the episodic mind.  
\bigskip

\subsection{Modeling the mimetic mind}
So far we have considered cognitive processes that occur in the episodic mind, which carries out appropriate responses to stimuli, but these responses are fixed. 
We now consider an additional process that is a distinguishing feature of the mimetic mind, which as mentioned earlier was physically larger than the episodic mind, and which Donald (1991) posited was capable of self-triggered recall and rehearsal. 

\begin{itemize}
\item[] {\bf Cognitive updating}:
In the process of {\it reflecting upon}, or thinking about an attended item $\ring{w} \in W_t$, we think about it in a new way or consider it appears from a different context or from another person's perspective, which we denote as $m \in M_t$. We say that $\ring{w} \in W_t$ is 
catalyzed by an item $m \in M_t$. This `reaction' updates the subject of thought, which is now denoted $\ring{w'} \in W_{t+\delta}$. We will refer to a single step such as this type of reflection process as {\em cognitive updating} and denote it by writing:
 $$\ring{w} \xrightarrow{m} \ring{w'}, \mbox{ and } \ring{w} \mapsto w.$$

% Since these processes are never differentiated in the model it doesn't seem appropriate to give them each bullet points here. LG
% \bigskip
% \begin{itemize}
% \item[] {\em Representational re-description}:
As an example, suppose you are thinking about a dog (this is $\ring{w}$) and you wonder what your mother would think of it (thus, $m$ is your mother's perspective, which plays the role of catalyst). Then $\ring{w'}$ is a new opinion of the dog that incorporates your mother's perspective. This example involves {\em representational re-description}, i.e., the modification or redescription of a MR of a dog.

\indent Abstract thought can also involve the {\em chaining}, or sequencing, of multiple MRs---such as representations for simple, single-step actions---into a more complex MR that involves multiple steps. It occasionally results in {\em concept combination}: the merger of two concepts into a more complex concept.

\end{itemize}

\begin{figure}[htb]
\centering
\includegraphics[scale=0.4]{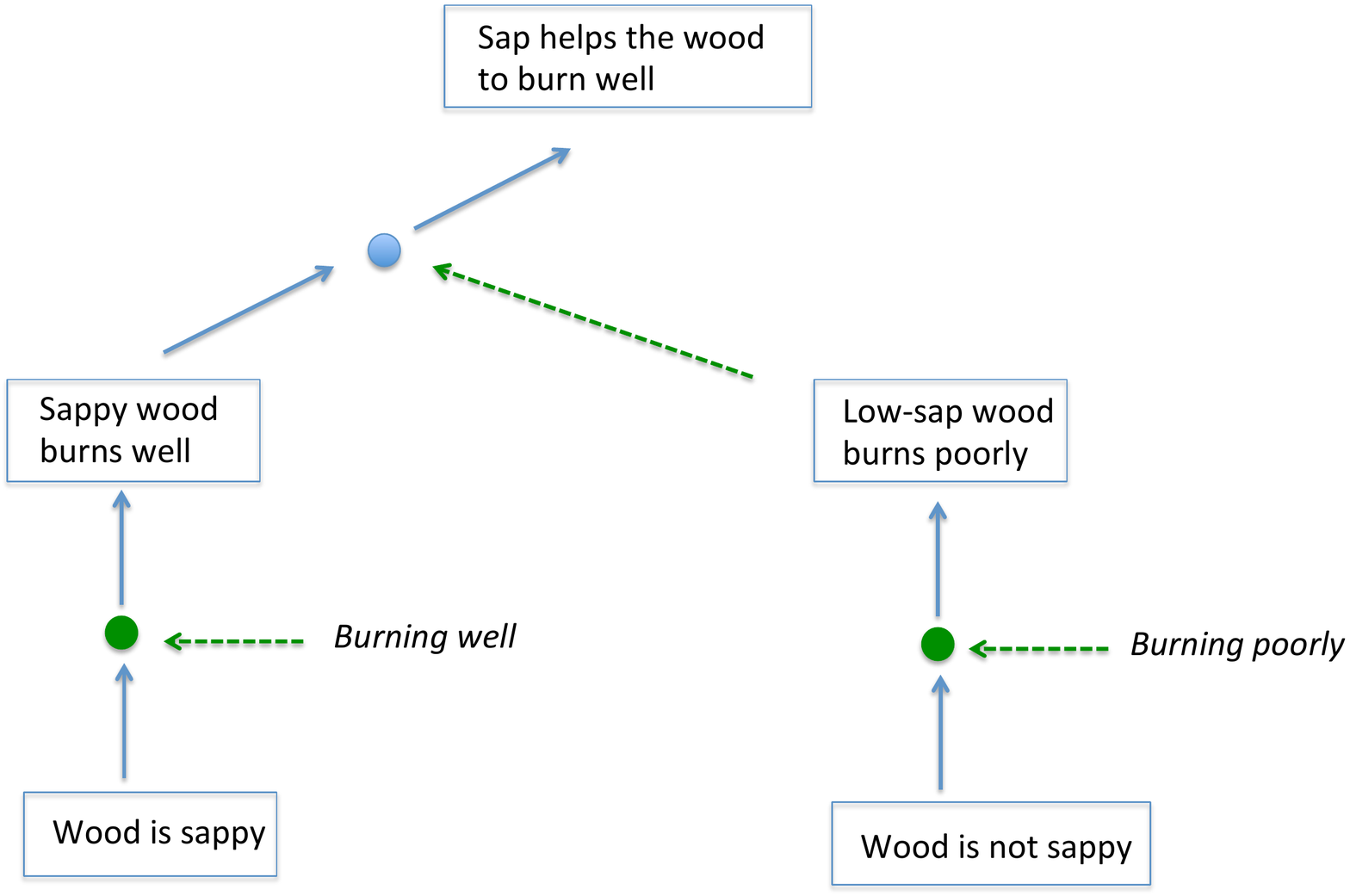}
\caption{The lower two reactions (green circles) correspond to `updating by stimulus' (i.e., $\ring{w} \xrightarrow{s} \ring{w'}$ where, for example, $\ring{w}=$ (wood is sappy), $s=$ (burning well), and $\ring{w'}$=(sappy wood burns well)). This kind of reaction is possible in either an episodic or mimetic mind. However, in the representational re-description reaction at the top of the figure (blue circle), a MR undergoes change in the absence of a stimulus; it is instead catalyzed by another MR. Representational re-description is the outcome of self-triggered recall, which is believed to be a distinguishing feature of a mimetic mind (Donald, 1991).}
\label{mimfig}
\end{figure}

Another example of representational re-description is illustrated by the top dashed arrow in Fig.~\ref{mimfig}.  Again, in order not to lose sight of the forest for the trees, we omit details of how causal relationships arise in cognition,  an active area of research in psychology and artificial intelligence largely carried out using Bayesian statistical models  \citep{good, tenen}.

% \end{itemize}

% \bigskip

% \item[] 
% \bigskip

% \item[] 

\mbox{}

Note that the key difference between this process and updating due to stimulus is the nature of the catalyst: here it is internal---i.e., an item in $M_t$---rather than external---i.e., a stimulus in $S_t$. 
In order to revise one's understanding of something, it was no longer necessary for something to happen in the physical world; this new understanding could arise due to `putting 2 and 2 together', or making more integrated use of thoughts and ideas encoded in memory. 

Notice also that there are various ways to model the fraction of mental representations that are close enough to the current subject of thought to generate a retrieval or reminding event. Under a binomial distribution, very few items are highly similar to any given item $m$, a great many are of intermediate similarity to $m$, and very few are extremely different from $m$. This  distribution widens as we allow for abstract categories, of which specific instances are members \citep{gab98}.

\section{Dynamics of cognition under the model}

\begin{figure}[htb]
\centering
\includegraphics[scale=0.85]{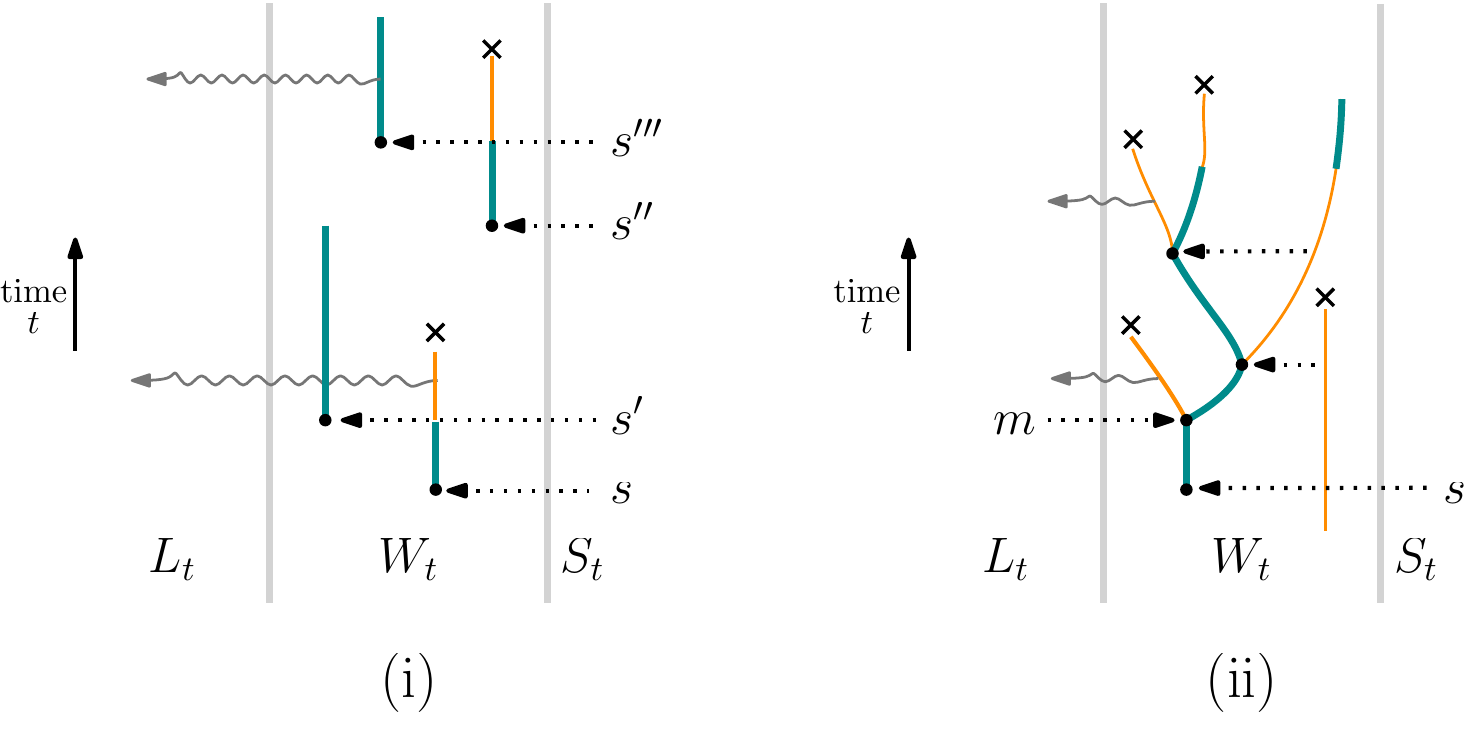}
\caption{In this figure, the attended item in $W_t$ is shown in solid green lines, with other items in $W_t$ as thin orange lines;  ${\bf \times}$ denotes that the item is no longer present in working memory.
(i) Item updating due to stimulus (the four reactions, with catalysis indicated by dotted arrows) together with encoding (an item from $W_t$ is cemented in $L_t$) is denoted by wiggly arrows). These do not allow for a cognitive catalytic processes (CCP) to form. 
(ii) The additional ability of items in $M_t$ to catalyze cognitive updating from $L_t$ (the lower dotted arrow) and  from within $W_t$ (the two uppermost dotted arrows). This leads to the formation of a CCP of size four. 
The disconnect in the solid green path near the top is an instance of a shift in attention to an item in working memory.}
\label{ccpfig}
\end{figure}

A {\em Cognitive Catalytic Process (CCP)} is a sequence of attended items $$\cC= \ring{w}_{t(1)}, \ring{w}_{t(2)}, \ldots, \ring{w}_{t(k)},$$ (where $\ring{w}_{t(i)} \in W_{t(i)}$, and where the $t(i)$ values are increasing) with the property that  each item $\ring{w}_{t(i)}$ after the first is generated from an earlier one by % either a cognitive updating reaction or an attention shift, including at least one cognitive 
a cognitive updating reaction.
In words, a CCP is a stream of thoughts, each of which builds on an earlier one, via its connection to (catalysis by) an item in long-term or working memory.
Newly generated MRs may subsequently be encoded in long-term memory and thus are available to catalyze further cognitive catalytic processes. 
We note that CCPs take shape in conjunction with drives and goals (though the details of how this works is beyond the scope of the current paper).
Fig.~\ref{ccpfig} provides a simple schematic example to illustrate the distinction between processes where CCPs are absent (i) and where they are present (ii). 
% removed second set of brackets around i & ii - lg

We suggest that by providing a mechanism whereby ideas can be combined, developed, enhanced, integrated with existing knowledge, and made available for further such processes, the emergences of CCPs can allow the development of a mimetic mind from a simpler episodic mind, regarded as a key step in the origin of cultural evolution.    The encoding of MRs arising from CCPs in long-term memory can then leads to a more integrated cognitive network (`conceptual closure') which we describe in Section~\ref{cograf}.

We now describe some generic features of the dynamics of CCPs and their emergence in the transition from an episodic to a mimetic mind. 
We focus on the impact of two parameters: the \emph{richness} of MRs (i.e., the detail with which items in memory are encoded) parameterized by the maximum number $N$ of properties of MRs, and their \emph{reactivity} (i.e., the extent to which features in a mental item trigger associations with other items), denoted $P$.
Here $N$ and $P$ can be viewed as the analogues of the maximum polymer length $M$ and the catalyzation probability $P$ (respectively) in Kauffman's OOL model from Section~\ref{early}.   
We will also describe how CCPs correspond to the autocatalytic network concepts of RAFs and CAFs that have been developed in origin-of-life research (Section~\ref{linky}).  

We begin by noting that whether or not a given MR in $M_t$ catalyzes a given cognitive updating reaction depends on numerous factors, such as how closely associated the items are in terms of shared properties, what stimuli are present, and what other MRs are active in working memory. The rate at which an item $m \in M_t$ catalyzes an attended item $\ring{w}$ in $W_t$ will be higher the more properties the two items share.

 Rather than trying to model the impact of increasing $N$ directly on the emergence of CCPs, we consider  the simpler case of increasing the average rate $\lambda$ at which items in $W_t$ and $L_t$ catalyze cognitive updating reactions (the rates within these two classes may differ, so $\lambda$ should be viewed as a scaling factor for both rates).  Note that $\lambda$ is an function of both $P$ and $N$.

We are particularly interested in understanding how the formation and persistence of CCPs  depends on this catalysis rate $\lambda$  and a possible transition that occurs when this catalysis rate increases, which could provide a feasible explanation for the transition from an episodic to a mimetic mind.  The following broad predictions, which can be easily derived in overly-simplified models (using techniques familiar from branching processes and random graph theory), are generic properties that would be expected to hold in  more specialized models. 

\begin{enumerate}
\item
When $\lambda$  is below a critical value, the dynamics of $\ring{w}_t, W_t$, and $M_t$ are essentially determined by the external stimuli $S_t$. 
This situation is characteristic of an episodic mind. If CCPs form at all, they do not persist, and therefore have negligible impact on the structure of the whole. Thus, items in memory remain essentially disconnected; they are activated in response to particular stimuli or situations, and evoke appropriate responses, but do not transform into an architecture that is continually revising its own structure by way of streams of thought. 

\item
As $\lambda$  increases towards a critical threshold, CCPs begin to form, and their size increases. This threshold depends on a sensitive interplay between $P$ and $L_t$, such that when long-term memory is denser, lower values of $P$ suffice. 

\item
The emergence of CCPs causes $M_t$ to grow at a faster rate than it would otherwise by generating a stream of thought that need not be related to current stimuli. Such streams of thought may be encoded in $M_t$ thereby providing a richer array of catalysts in $L_t$ for future cognitive updating reactions (and thereby CCPs) and so generating a positive feedback process. This situation is characteristic of the mimetic mind.
%revised this passage a bit - lg

\end{enumerate}

\section{Cognitive RAFs: The mind as an autocatalytic network}
\label{linky}

The model that we have described can be viewed as an instance of an autocatalytic set in an abstract reactive network, as described in Section ~\ref{rafback}, in which the transition from episodic to mimetic mind corresponds to the emergence of an RAF set. To describe this more formally, 
consider the myriad of ways that $M_t$ could develop from $M_0$ (at conception) to its state at some particular time $t$. 
More precisely, for a fixed time $t>0$, consider the following  catalytic reaction system $\cQ = (X, \cR,  C, F)$, where: 
\begin{itemize}
\item $F = F_t = \left( \bigcup_{t' \leq t} S_{t'} \right) \cup M_0$ (this is what the external environment provides (stimuli);
\item $X = \left(\bigcup_{t' \leq t} M_{t'}\right) \cup F_t$ (this is the set of all mental representations that have been present in the mind from conception up to some time $t$, together with $F_t$). 
\item $\cR$ is the set of all the updating reactions that can potentially take place from conception up to  time $t$;
\item $C$ is all the catalysis assignments that are potentially possible.
\end{itemize}
$\cR$ and $C$ are not prescribed in advance. Because $C$ includes remindings and associations on the basis of, not just a single shared property, but also on the basis of multiple shared properties, MRs can develop in a potentially unlimited number of directions through interactions with other MRs. Nevertheless, it makes perfect mathematical sense to talk about $\cR$ and $C$ as sets. 
%reworded this slightly -L
The justification of the following result is provided in the Appendix, in which we assume that $t$ is large enough that updating reactions have commenced, and that the rate at which stimuli occur is bounded.
\begin{proposition}
\label{mainpro}
$\cQ$ contains a RAF that increases in size with $t$ (namely the set of updating reactions that actually do occur between time 0 and $t$). Moreover, while  $\lambda$ is below a critical threshold, CCPs in this RAF are short and few in number, but when   $\lambda$  exceeds this threshold, CCPs become more frequent,  persistent and complex.
\end{proposition}
We will refer to the particular RAF described in  Proposition~\ref{mainpro} as the {\em cognitive RAF}. It has, in fact, the stronger property of being a CAF, as defined in \cite{mos}. The significance of a transition from linear to super-linear growth in Proposition~\ref{mainpro} is in providing a mechanism for generating densely linked connections in the mimetic mind. This is described in more detail in the next section. 

This formal connection  between the RAF structure of (i) our simple model of the mind and (ii) a model that has been used to understand the transition to self-sustaining autocatalytic life in biochemistry,  may be helpful in subsequent work. This is because efficient (polynomial-time) algorithms exist for detecting RAFs (and CAFs) in catalytic reaction systems in general, and for studying their organisation and structure. (For further details, see \cite{hor16} and the references therein). 

\subsection{Cognitive RAFs and conceptual closure}
\label{cograf}
Our simple model provides a mechanism by which items involved in the cognitive RAF (reactants, products, and catalysts) and in the sequences (streams of thought) that form CCPs  present in this cognitive RAF
can give rise (via the encoding process) to increasingly interlinked associations between mental representations in long-term memory.  
%yyy
Formally, we  say that a set ${\mathcal C}$ of items in $L_t$ (i.e., long-term memory) is a {\em conceptually closed set} 
if, for any two items $m_i, m_j$  in ${\mathcal C}$, there is a possible sequence of cognitive updating reactions that (if activated) 
can relate $m_i$ to $m_j$, and such that each reaction in that sequence has a catalyst that is also an item in ${\mathcal C}$. 
In this definition, saying that the sequence of reactions {\em relates} $m_i$ to $m_j$ means that for each reaction in the sequence, the reactant and product are
associated (here `associated' is as defined in the paragraphs before Section~\ref{forging}).

As $N$
increases in the transition from episodic mind to mimetic mind, $L_t$ begins to increase more rapidly (via Proposition~\ref{mainpro}) and  
$P$  becomes tuned to match $N$ such that reminding events are neither too frequent, such that the network is super-critical (an `everything reminds you of everything' situation), nor too infrequent, such that the network is subcritical (a `nothing reminds you of anything' situation), as discussed in Section 3.
Thus, conceptually closed sets of increasing size and complexity can form.  

\section{Conclusions}

We suggest that it was not a change to any particular brain area that enabled the threshold to cumulative cultural evolution to be crossed, but a change to how the brain functions as a  whole, and this change can be articulated using a mathematical model. 

It has been proposed that cultural evolution was made possible by a cognitive transition brought about by onset of the capacity for self-triggered recall and rehearsal. However, self-triggered recall requires that concepts and ideas be accessible to one another (i.e., that they collectively constitute an integrated structure). We suggest that, much as models of self-sustaining, autocatalytic networks have been useful for understanding how the the origin of life, and thus biological evolution, could have come about, they are also useful for understanding how the the origin of the kind of cognitive structure that makes cultural evolution possible could have come about. Mental representations (such as memories, concepts, and schemas) play the role of `reactants' and `catalysts', and relationships amongst them (such as associations, remindings, and causal relationships) are the `reactions'. In the pre-cultural `episodic' mind, such reactions are catalyzed only by external stimuli. As cranial capacity increases, representations become richer (more features or properties are encoded), and thus 
reactions become more plentiful, leading to streams of thought. Streams of thought cause the reaction network to become even denser. Eventually, it becomes almost inevitable that a percolation threshold is surpassed, and collectively the representations form an integrated autocatalytic set. At this point, the mind can combine representations and adapt them to specific needs and situations, and thereby become a contributor to culture. We posit that an interacting population of such minds is capable of cumulatively creative cultural evolution.

Our model provides a means of differentiating between the episodic mind of {\it Homo habilis}, the mimetic mind of {\it Homo erectus}, and the mind of a young child. The proposed similarities and differences amongst them are summarized in Table 2.

\begin{table}[ht]
  \begin{center}
  \begin{tabular}{@{} clccccrrr @{}}
    \hline \hline 
   \bf{Variable} & \bf{Variable} & \bf{Homo} & \bf{Homo} & \bf{Young} \\
   \bf{(Symbol)} & \bf{(In Words)} & \bf{habilis} & \bf{erectus} & \bf{Child} \\
      \hline \hline 
     $N$ & MR richness & Low & High & High \\
    \hline 
    $X$ & Set of existing MRs & Large & Large & Small \\
    \hline 
    $P$  & Reactivity & Fixed & Tuned to $N$ & Tuned to $N$ \\
         \hline 
 $\lambda$  & Catalysis rate & Small & Larger & Small \\
     \hline
   $W_t$ & Working memory & Small & Larger & Small \\
        \hline
 $L_t$  & Long-term memory & Small & Much larger & Small \\
        \hline
$M_t$  & Memory ($W_t$ and/or $L_t$) & Small & Much larger & Small \\
        \hline
   $CCP$s & Streams of thought & Absent & Present & Absent \\
   \hline
   ${\mathcal C}$  & Conceptual Closure & Absent & Present & Absent \\     
    \hline \hline 
  \end{tabular}
  \end{center}
\vspace{1mm}{\refstepcounter{table}\label{tab:comparison}Table \arabic{table}. Summary comparison of the episodic mind of {\it Homo habilis}, the mimetic mind of {\it Homo erectus}, and the mind of a young child.}
\end{table}

We suggest that in the mind of a developing child, MRs are sufficiently rich, and the catalysis rate is sufficiently high, but memory is not yet packed densely with enough MRs for CCPs to occur. As more MRs are encoded in long-term memory, the effective rate of cognitive updating increases as items get encoded into long-term memory (the more items there, the more likely one is to catalyze a cognitive updating reaction).  This, in turn, increases working memory, which also eventually increases the rate of cognitive updating reactions from within working memory.\footnote{This is supported by findings that measures of performance on tests of working memory increase continuously between early childhood and adolescence \citep{gat}.}
In short, 
while the OOC is attributed to an increase in $N$ and corresponding fine-tuning of $P$, in the developmental transition of a young child to a contributing member of culture, $N$ and $P$ are sufficiently high but $X$ is not sufficiently dense for the formation of CCPs.

While our model is individual-based, it also allows for societal interactions and development. More precisely, if we have a collection of minds (society), then an item $m \in M_t$ in one individual can (through speech, gesture, or action) be regarded as a stimulus $s$ for another individual, and thereby provide an `updating by stimuli' reaction in that individual. Thus the collection of minds (together with other stimuli from the environment) provides a higher-level network structure.

There are several caveats and limitations to this work. 
Firstly, we have not dealt with the problems that have arisen in psychology and artificial intelligence in trying to deal with mental representations, reasoning and inference, creativity, and language understanding. These challenges are not the subject of this paper. Nevertheless, we believe that by introducing an architecture conducive to self-organized emergent cognitive complexity, the proposed framework has the potential to facilitate such efforts.

Also, 
in this paper we have not provided a mechanism that accounts for awareness, though one of us has proposed such a mechanism elsewhere (\cite{gab02b}). Nor have we provided a mechanism for subconscious processing within working memory, although one of us has published on this extensively elsewhere (\cite{gab02, gab10, gabinpress, gabran13}). In a subsequent paper, we will aim to show how implicit processing fit into this model.

% we have not provided a mechanism that accounts for awareness, nor have we provided an explicit mechanism for subconscious processing within working memory.}

Indeed, the model proposed here is quite general and schematic.
To make it precise enough to allow a detailed mathematical analysis requires specifying a large number of assumptions and parameter choices, estimated from empirical studies. Since the goal of the present paper is merely to show that, for a range of reasonable values, the kind of cognitive reorganization that we propose made cultural evolution possible is likely to occur, and leads to testable predictions. Rather than exploring any particular choice here---a topic that we wish to pursue in future work---our approach is to consider generic properties of the simple and general model described. 

Other fruitful arenas for future research would be to more fully explore how transitions in the network structure map onto cognitive developmental stages, or how different ways of achieving a closure structure map onto personality differences. We might speculatively suggest that the fact that there are different ways of satisfying the criterion (for example, very high reactivity with a medium number of episodes, versus a very high number of episodes and medium reactivity) could form the basis of fascinating personality test. A person with high reactivity might be likely to understand things in terms of analogies and metaphors and make decisions in a context-dependent way, whereas a person with a high number of episodes would be more likely to understand things in terms of their large repertoire of cultural teachings and make decisions according to how its been done in the past as opposed to taking context-specific factors into account.

We conclude by suggesting that the common mathematical approach of two superficially different evolution processes (the origin of life and the origin of culture) depends on a certain kind of deep abstract structure, which has also recently been identified in other fields, such a ecology \citep{gatt} and economics \citep{wim}. This may prove useful for studying emergent processes in other fields. 

% If you think the above paragraph is too corny we can leave it out :)

\section{Acknowledgments}
We thank an anonymous reviewer for a number of helpful comments and suggestions.
This work was supported by a grant (62R06523) from the Natural Sciences and Engineering Research Council of Canada.

\section*{References}
\bibliographystyle{model2-names}
\bibliography{bibtex.bib}

\begin{thebibliography}{50}
\expandafter\ifx\csname natexlab\endcsname\relax\def\natexlab#1{#1}\fi
\expandafter\ifx\csname url\endcsname\relax
  \def\url#1{\texttt{#1}}\fi
\expandafter\ifx\csname urlprefix\endcsname\relax\def\urlprefix{URL }\fi
\providecommand{\eprint}[2][]{\url{#2}}
\providecommand{\bibinfo}[2]{#2}
\ifx\xfnm\relax \def\xfnm[#1]{\unskip,\space#1}\fi
%Type = Incollection
\bibitem[{Aiello(1996)}]{aie96}
\bibinfo{author}{Aiello, L.C.}, \bibinfo{year}{1996}.
\newblock \bibinfo{title}{Hominine pre-adaptations for language and cognition},
  in: \bibinfo{editor}{Mellars, P.}, \bibinfo{editor}{Gibson, K.} (Eds.),
  \bibinfo{booktitle}{Modeling the early human mind}.
  \bibinfo{publisher}{McDonald Institute Monographs},
  \bibinfo{address}{Cambridge, UK}, pp. \bibinfo{pages}{89--99}.
%Type = Book
\bibitem[{Cabell and Valsiner(2014)}]{cat14}
\bibinfo{author}{Cabell, K.R.}, \bibinfo{author}{Valsiner, J.},
  \bibinfo{year}{2014}.
\newblock \bibinfo{title}{The catalyzing mind: {B}eyond models of causality}.
\newblock Annals of Theoretical Psychology 11, \bibinfo{publisher}{Springer}.
%Type = Book
\bibitem[{Chomsky(2012)}]{cho12}
\bibinfo{author}{Chomsky, N.}, \bibinfo{year}{2012}.
\newblock \bibinfo{title}{The science of language}.
\newblock \bibinfo{publisher}{Cambridge University Press}.
%Type = Book
\bibitem[{Corballis(2011)}]{cor11}
\bibinfo{author}{Corballis, M.C.}, \bibinfo{year}{2011}.
\newblock \bibinfo{title}{The recursive mind: the origins of human language,
  thought and civilization}.
\newblock \bibinfo{publisher}{Princeton University Press},
  \bibinfo{address}{Princeton, NJ}.
%Type = Book
\bibitem[{Donald(1991)}]{don91}
\bibinfo{author}{Donald, M.}, \bibinfo{year}{1991}.
\newblock \bibinfo{title}{Origins of the modern mind: Three stages in the
  evolution of culture and cognition}.
\newblock \bibinfo{publisher}{Harvard University Press},
  \bibinfo{address}{Cambridge, MA}.
%Type = Article
\bibitem[{Erd{\"o}s and R{\'e}nyi(1960)}]{erdosrenyi1960}
\bibinfo{author}{Erd{\"o}s, P.}, \bibinfo{author}{R{\'e}nyi, A.},
  \bibinfo{year}{1960}.
\newblock \bibinfo{title}{On the evolution of random graphs}.
\newblock \bibinfo{journal}{Publications of the Mathematical Institute of the
  Hungarian Academy of Sciences} \bibinfo{volume}{5}, \bibinfo{pages}{17--61}.
%Type = Article
\bibitem[{Farmer et~al.(1986)Farmer, Kauffman and Packard}]{farmeretal1986}
\bibinfo{author}{Farmer, J.D.}, \bibinfo{author}{Kauffman, S.A.},
  \bibinfo{author}{Packard, N.H.}, \bibinfo{year}{1986}.
\newblock \bibinfo{title}{Autocatalytic replication of polymers}.
\newblock \bibinfo{journal}{Physica D: Nonlinear Phenomena}
  \bibinfo{volume}{22}, \bibinfo{pages}{50--67}.
%Type = Article
\bibitem[{Gabora(1998)}]{gab98}
\bibinfo{author}{Gabora, L.}, \bibinfo{year}{1998}.
\newblock \bibinfo{title}{Autocatalytic closure in a cognitive system: A
  tentative scenario for the origin of culture}.
\newblock \bibinfo{journal}{Psycoloquy} \bibinfo{volume}{9}, \bibinfo{pages}{67
  [adap--org/9901002]}.
%Type = Incollection
\bibitem[{Gabora(2000)}]{gab00}
\bibinfo{author}{Gabora, L.}, \bibinfo{year}{2000}.
\newblock \bibinfo{title}{Conceptual closure: {H}ow memories are woven into an
  interconnected worldview.}, in: \bibinfo{editor}{Van~de Vijver, G.},
  \bibinfo{editor}{Chandler, J.} (Eds.), \bibinfo{booktitle}{Closure: Emergent
  Organizations and their Dynamics}. \bibinfo{publisher}{Annals of the New York
  Academy of Sciences}. \bibinfo{number}{901}, pp. \bibinfo{pages}{42--53}.
%Type = Phdthesis
\bibitem[{Gabora(2001)}]{gab01}
\bibinfo{author}{Gabora, L.}, \bibinfo{year}{2001}.
\newblock \bibinfo{title}{Cognitive mechanisms underlying the origin and
  evolution of culture}.
\newblock \bibinfo{type}{Doctoral dissertation}. Free University of Brussels,
  Belgium.
%Type = Article
\bibitem[{Gabora(2002a)}]{gab02b}
\bibinfo{author}{Gabora, L.}, \bibinfo{year}{2002}a.
\newblock \bibinfo{title}{Amplifying phenomenal information: Toward a
  fundamental theory of consciousness.}
\newblock \bibinfo{journal}{Journal of Consciousness Studies}
  \bibinfo{volume}{9}, \bibinfo{pages}{3--29}.
%Type = Incollection
\bibitem[{Gabora(2002b)}]{gab02}
\bibinfo{author}{Gabora, L.}, \bibinfo{year}{2002}b.
\newblock \bibinfo{title}{Cognitive mechanisms underlying the creative
  process.}, in: \bibinfo{editor}{T., H.}, \bibinfo{editor}{Kavanagh, T.}
  (Eds.), \bibinfo{booktitle}{Proceedings of the Fourth International
  Conference on Creativity and Cognition}. \bibinfo{publisher}{MIT Press},
  \bibinfo{address}{Cambridge, MA}, pp. \bibinfo{pages}{126--133}.
%Type = Article
\bibitem[{Gabora(2006)}]{gab06}
\bibinfo{author}{Gabora, L.}, \bibinfo{year}{2006}.
\newblock \bibinfo{title}{Self-other organization: Why early life did not
  evolve through natural selection}.
\newblock \bibinfo{journal}{Journal of Theoretical Biology}
  \bibinfo{volume}{241}, \bibinfo{pages}{443--450}.
%Type = Article
\bibitem[{Gabora(2010)}]{gab10}
\bibinfo{author}{Gabora, L.}, \bibinfo{year}{2010}.
\newblock \bibinfo{title}{Revenge of the `neurds': Characterizing creative
  thought in terms of the structure and dynamics of human memory.}
\newblock \bibinfo{journal}{22} , \bibinfo{pages}{1--13}.
%Type = Article
\bibitem[{Gabora(2013)}]{gab13}
\bibinfo{author}{Gabora, L.}, \bibinfo{year}{2013}.
\newblock \bibinfo{title}{An evolutionary framework for culture: Selectionism
  versus communal exchange}.
\newblock \bibinfo{journal}{Physics of Life Reviews} \bibinfo{volume}{10},
  \bibinfo{pages}{117--145}.
%Type = Incollection
\bibitem[{Gabora(2018a)}]{gab-press}
\bibinfo{author}{Gabora, L.}, \bibinfo{year}{2018}a.
\newblock \bibinfo{title}{The creative process of cultural evolution}, in:
  \bibinfo{editor}{Leung, A.} (Ed.), \bibinfo{booktitle}{Handbook of Culture
  and creativity: Basic processes and applied innovations}.
  \bibinfo{publisher}{Oxford University Press}.
%Type = Incollection
\bibitem[{Gabora(2018b)}]{gabinpress}
\bibinfo{author}{Gabora, L.}, \bibinfo{year}{2018}b.
\newblock \bibinfo{title}{How insight emerges in distributed,
  content-addressable memory}, in: \bibinfo{editor}{Vartanian, O.},
  \bibinfo{editor}{Jung, J.} (Eds.), \bibinfo{booktitle}{The Cambridge Handbook
  of the Neuroscience of Creativity}. \bibinfo{publisher}{MIT Press},
  \bibinfo{address}{Cambridge, MA}.
%Type = Incollection
\bibitem[{Gabora and Ranjan(2013)}]{gabran13}
\bibinfo{author}{Gabora, L.}, \bibinfo{author}{Ranjan, A.},
  \bibinfo{year}{2013}.
\newblock \bibinfo{title}{How insight emerges in distributed,
  content-addressable memory}, in: \bibinfo{editor}{Bristol, A.},
  \bibinfo{editor}{Vartanian, O.}, \bibinfo{editor}{Kaufman, J.} (Eds.),
  \bibinfo{booktitle}{The neuroscience of creativity}. \bibinfo{publisher}{MIT
  Press}, \bibinfo{address}{Cambridge, MA}, pp. \bibinfo{pages}{19--43}.
%Type = Unpublished
\bibitem[{Gabora and Smith(2017)}]{gab-prep}
\bibinfo{author}{Gabora, L.}, \bibinfo{author}{Smith, L.},
  \bibinfo{year}{2017}.
\newblock \bibinfo{title}{Two cognitive transitions underlying the capacity for
  cultural evolution}.
\newblock \bibinfo{note}{(submitted.)}.
%Type = Article
\bibitem[{Gathercole et~al.((2004)Gathercole, Pickering, Ambridge and
  Wearing}]{gat}
\bibinfo{author}{Gathercole, S.E.}, \bibinfo{author}{Pickering, S.J.},
  \bibinfo{author}{Ambridge, B.}, \bibinfo{author}{Wearing, H.},
  \bibinfo{year}{(2004}.
\newblock \bibinfo{title}{The structure of working memory from 4 to 15 years of
  age}.
\newblock \bibinfo{journal}{Developmental Psychology} \bibinfo{volume}{40},
  \bibinfo{pages}{177--190}.
\newblock \bibinfo{note}{Doi:10.1037/0012-1649.40.2.177. PMID 14979759.}
%Type = Article
\bibitem[{Gatti et~al.(2017)Gatti, Hordijk and Kauffman}]{gatt}
\bibinfo{author}{Gatti, R.C.}, \bibinfo{author}{Hordijk, W.},
  \bibinfo{author}{Kauffman, S.}, \bibinfo{year}{2017}.
\newblock \bibinfo{title}{Biodiversity is autocatalytic}.
\newblock \bibinfo{journal}{Ecological modelling} \bibinfo{volume}{346},
  \bibinfo{pages}{70--76}.
%Type = Article
\bibitem[{Gibbons(1998)}]{gib98}
\bibinfo{author}{Gibbons, A.}, \bibinfo{year}{1998}.
\newblock \bibinfo{title}{Ancient island tools suggest \textit{{H}omo erectus}
  was a seafarer}.
\newblock \bibinfo{journal}{Science} \bibinfo{volume}{297},
  \bibinfo{pages}{1635--1637}.
%Type = Article
\bibitem[{Goodman et~al.(2011)Goodman, Ullmand and Tenenbaum}]{good}
\bibinfo{author}{Goodman, N.D.}, \bibinfo{author}{Ullmand, T.D.},
  \bibinfo{author}{Tenenbaum, J.B.}, \bibinfo{year}{2011}.
\newblock \bibinfo{title}{Learning a theory of causality}.
\newblock \bibinfo{journal}{Psychological review} \bibinfo{volume}{118},
  \bibinfo{pages}{110}.
%Type = Article
\bibitem[{Goren-Inbar et~al.(2004)Goren-Inbar, Alperson, Kiselv, Simchoni,
  Melamed, Ben-Nun and Werker}]{gor04}
\bibinfo{author}{Goren-Inbar, N.}, \bibinfo{author}{Alperson, N.},
  \bibinfo{author}{Kiselv, M.E.}, \bibinfo{author}{Simchoni, O.},
  \bibinfo{author}{Melamed, Y.}, \bibinfo{author}{Ben-Nun, A.},
  \bibinfo{author}{Werker, E.}, \bibinfo{year}{2004}.
\newblock \bibinfo{title}{Evidence of hominin control of fire at {G}esher
  {B}enot {Y}a'aqov, {I}srael}.
\newblock \bibinfo{journal}{Science} \bibinfo{volume}{304},
  \bibinfo{pages}{725--727}.
%Type = Incollection
\bibitem[{Haidle(2009)}]{hai09}
\bibinfo{author}{Haidle, M.N.}, \bibinfo{year}{2009}.
\newblock \bibinfo{title}{How to think a simple spear}, in:
  \bibinfo{editor}{deBaune, S.A.}, \bibinfo{editor}{Coolidge, F.L.},
  \bibinfo{editor}{Wynn, T.} (Eds.), \bibinfo{booktitle}{Cognitive archaeology
  and human evolution}. \bibinfo{publisher}{Cambridge University Press},
  \bibinfo{address}{Cambridge, MA}, pp. \bibinfo{pages}{57--73}.
%Type = Article
\bibitem[{Harmand et~al.(2015)Harmand, Lewis, Feibel, Lepre, Prat, Lenoble,
  Bo\"{e}s, Quinn, Brenet, Arroyo, Taylor, Cl\'{e}ment, Daver, Brugal, Leakey,
  Mortlock, Wright, Lokorodi, Kirwa, Kent and Roche}]{har15}
\bibinfo{author}{Harmand, S.}, \bibinfo{author}{Lewis, J.E.},
  \bibinfo{author}{Feibel, C.S.}, \bibinfo{author}{Lepre, C.J.},
  \bibinfo{author}{Prat, S.}, \bibinfo{author}{Lenoble, A.},
  \bibinfo{author}{Bo\"{e}s, X.}, \bibinfo{author}{Quinn, R.L.},
  \bibinfo{author}{Brenet, M.}, \bibinfo{author}{Arroyo, A.},
  \bibinfo{author}{Taylor, N.}, \bibinfo{author}{Cl\'{e}ment, S.},
  \bibinfo{author}{Daver, G.}, \bibinfo{author}{Brugal, J.P.},
  \bibinfo{author}{Leakey, L.}, \bibinfo{author}{Mortlock, R.A.},
  \bibinfo{author}{Wright, J.D.}, \bibinfo{author}{Lokorodi, S.},
  \bibinfo{author}{Kirwa, C.}, \bibinfo{author}{Kent, D.V.},
  \bibinfo{author}{Roche, H.}, \bibinfo{year}{2015}.
\newblock \bibinfo{title}{3.3-million-year-old stone tools from {L}omekwi 3,
  {W}est {T}urkana, {K}enya}.
\newblock \bibinfo{journal}{Nature} \bibinfo{volume}{521},
  \bibinfo{pages}{310--315}.
%Type = Article
\bibitem[{Hauser et~al.(2002)Hauser, Chomsky and Fitch}]{hau02}
\bibinfo{author}{Hauser, M.D.}, \bibinfo{author}{Chomsky, N.},
  \bibinfo{author}{Fitch, W.T.}, \bibinfo{year}{2002}.
\newblock \bibinfo{title}{The faculty of language: What is it, who has it and
  how did it evolve?}
\newblock \bibinfo{journal}{Science} \bibinfo{volume}{298},
  \bibinfo{pages}{1569--1579}.
%Type = Article
\bibitem[{Hordijk(2013)}]{wim}
\bibinfo{author}{Hordijk, W.}, \bibinfo{year}{2013}.
\newblock \bibinfo{title}{From the origin of life to the economy}.
\newblock \bibinfo{journal}{BioScience} \bibinfo{volume}{63},
  \bibinfo{pages}{877--881}.
%Type = Article
\bibitem[{Hordijk et~al.(2010)Hordijk, Hein and Steel}]{horetal10}
\bibinfo{author}{Hordijk, W.}, \bibinfo{author}{Hein, J.},
  \bibinfo{author}{Steel, M.}, \bibinfo{year}{2010}.
\newblock \bibinfo{title}{Autocatalytic sets and the origin of life}.
\newblock \bibinfo{journal}{Entropy} \bibinfo{volume}{12},
  \bibinfo{pages}{1733--1742}.
%Type = Article
\bibitem[{Hordijk et~al.(2011)Hordijk, Kauffman and Steel}]{horetal11}
\bibinfo{author}{Hordijk, W.}, \bibinfo{author}{Kauffman, S.A.},
  \bibinfo{author}{Steel, M.}, \bibinfo{year}{2011}.
\newblock \bibinfo{title}{Required levels of catalysis for emergence of
  autocatalytic sets in models of chemical reaction systems}.
\newblock \bibinfo{journal}{International Journal of Molecular Sciences}
  \bibinfo{volume}{12}, \bibinfo{pages}{3085--3101}.
%Type = Article
\bibitem[{Hordijk and Steel(2004)}]{horsteel04}
\bibinfo{author}{Hordijk, W.}, \bibinfo{author}{Steel, M.},
  \bibinfo{year}{2004}.
\newblock \bibinfo{title}{Detecting autocatalytic, self-sustaining sets in
  chemical reaction systems}.
\newblock \bibinfo{journal}{Journal of Theoretical Biology}
  \bibinfo{volume}{227}, \bibinfo{pages}{451--461}.
%Type = Article
\bibitem[{Hordijk and Steel(2012)}]{horsteel12}
\bibinfo{author}{Hordijk, W.}, \bibinfo{author}{Steel, M.},
  \bibinfo{year}{2012}.
\newblock \bibinfo{title}{Predicting template-based catalysis rates in a simple
  catalytic reaction model}.
\newblock \bibinfo{journal}{Journal of Theoretical Biology}
  \bibinfo{volume}{295}, \bibinfo{pages}{132--138}.
%Type = Article
\bibitem[{Hordijk and Steel(2016)}]{hor16}
\bibinfo{author}{Hordijk, W.}, \bibinfo{author}{Steel, M.},
  \bibinfo{year}{2016}.
\newblock \bibinfo{title}{Chasing the tail: The emergence of autocatalytic
  networks}.
\newblock \bibinfo{journal}{Biosystems} \bibinfo{volume}{152},
  \bibinfo{pages}{1--10}.
%Type = Article
\bibitem[{Kauffman(1986)}]{kau2}
\bibinfo{author}{Kauffman, S.A.}, \bibinfo{year}{1986}.
\newblock \bibinfo{title}{Autocatalytic sets of proteins}.
\newblock \bibinfo{journal}{Journal of Thoeretical Biology}
  \bibinfo{volume}{119}, \bibinfo{pages}{1--24}.
%Type = Book
\bibitem[{Kauffman(1993)}]{kau1}
\bibinfo{author}{Kauffman, S.A.}, \bibinfo{year}{1993}.
\newblock \bibinfo{title}{The origins of order}.
\newblock \bibinfo{publisher}{Oxford University Press}.
%Type = Article
\bibitem[{Lepre et~al.(2011)Lepre, Roche, Kent, Harmand, Quinn, Brugal, Texier,
  Lenoble and Feibel}]{lep11}
\bibinfo{author}{Lepre, C.J.}, \bibinfo{author}{Roche, H.},
  \bibinfo{author}{Kent, D.V.}, \bibinfo{author}{Harmand, S.},
  \bibinfo{author}{Quinn, R.L.}, \bibinfo{author}{Brugal, J.P.},
  \bibinfo{author}{Texier, P.J.}, \bibinfo{author}{Lenoble, A.},
  \bibinfo{author}{Feibel, C.S.}, \bibinfo{year}{2011}.
\newblock \bibinfo{title}{An earlier origin for the {A}cheulian}.
\newblock \bibinfo{journal}{Nature} \bibinfo{volume}{477},
  \bibinfo{pages}{82--85}.
%Type = Incollection
\bibitem[{Mania and Mania(2005)}]{man05}
\bibinfo{author}{Mania, D.}, \bibinfo{author}{Mania, U.}, \bibinfo{year}{2005}.
\newblock \bibinfo{title}{The natural and socio-cultural environment of
  \textit{{H}omo erectus} at {B}ilzingsleben, {G}ermany}, in:
  \bibinfo{editor}{Gamble, C.} (Ed.), \bibinfo{booktitle}{The Individual
  Hominid in context: Archaeological investigations of Lower and Middle
  Palaeolithic landscapes, locales, and artefacts}.
  \bibinfo{publisher}{Routledge}, \bibinfo{address}{New York}, pp.
  \bibinfo{pages}{98--114}.
%Type = Book
\bibitem[{Mithen(1998)}]{mit98}
\bibinfo{author}{Mithen, S.}, \bibinfo{year}{1998}.
\newblock \bibinfo{title}{Creativity in human evolution and prehistory}.
\newblock \bibinfo{publisher}{Routledge}, \bibinfo{address}{London, UK}.
%Type = Article
\bibitem[{Mossel and Steel(2005)}]{mos}
\bibinfo{author}{Mossel, E.}, \bibinfo{author}{Steel, M.},
  \bibinfo{year}{2005}.
\newblock \bibinfo{title}{Random biochemical networks and the probability of
  self-sustaining autocatalysis}.
\newblock \bibinfo{journal}{Journal of Theoretical Biology}
  \bibinfo{volume}{233}, \bibinfo{pages}{327--336}.
%Type = Book
\bibitem[{Moutsou(2014)}]{mou14}
\bibinfo{author}{Moutsou, T.}, \bibinfo{year}{2014}.
\newblock \bibinfo{title}{The obsidian evidence for the scale of social life
  during the Palaeolithic}.
\newblock British Archaeological Reports International Series \#2613,
  \bibinfo{publisher}{Oxford}.
%Type = Article
\bibitem[{New and Pohorille(2000)}]{new00}
\bibinfo{author}{New, M.H.}, \bibinfo{author}{Pohorille, A.},
  \bibinfo{year}{2000}.
\newblock \bibinfo{title}{An inherited efficiencies model of non-genomic
  evolution}.
\newblock \bibinfo{journal}{Simulation Practice Theory} \bibinfo{volume}{8},
  \bibinfo{pages}{99--108}.
%Type = Article
\bibitem[{Parfitt et~al.(2010)Parfitt, Ashton, Lewis, Abel, Coope, Field, Gale,
  Hoare, Larkin, Lewis, Karloukivski, Maher, Peglar, Preece, Whittaker and
  Stringer}]{par10}
\bibinfo{author}{Parfitt, S.}, \bibinfo{author}{Ashton, N.M.},
  \bibinfo{author}{Lewis, S.G.}, \bibinfo{author}{Abel, R.L.},
  \bibinfo{author}{Coope, G.R.}, \bibinfo{author}{Field, M.H.},
  \bibinfo{author}{Gale, R.}, \bibinfo{author}{Hoare, P.G.},
  \bibinfo{author}{Larkin, N.R.}, \bibinfo{author}{Lewis, M.D.},
  \bibinfo{author}{Karloukivski, V.}, \bibinfo{author}{Maher, B.A.},
  \bibinfo{author}{Peglar, S.M.}, \bibinfo{author}{Preece, R.C.},
  \bibinfo{author}{Whittaker, J.F.}, \bibinfo{author}{Stringer, C.B.},
  \bibinfo{year}{2010}.
\newblock \bibinfo{title}{Early {P}leistocene human occupation at the edge of
  the boreal zone in northwest {E}urope}.
\newblock \bibinfo{journal}{Nature} \bibinfo{volume}{466},
  \bibinfo{pages}{229--233}.
%Type = Article
\bibitem[{Penn et~al.(2008)Penn, Holyoak and Povinelli}]{pen08}
\bibinfo{author}{Penn, D.}, \bibinfo{author}{Holyoak, K.},
  \bibinfo{author}{Povinelli, D.}, \bibinfo{year}{2008}.
\newblock \bibinfo{title}{Darwin's mistake: {E}xplaining the discontinuity
  between human and nonhuman minds}.
\newblock \bibinfo{journal}{Behavioral and Brain Sciences}
  \bibinfo{volume}{31}, \bibinfo{pages}{109--178}.
%Type = Article
\bibitem[{Plummer(2004)}]{plu04}
\bibinfo{author}{Plummer, T.}, \bibinfo{year}{2004}.
\newblock \bibinfo{title}{Flaked stones and old bones: Biological and cultural
  evolution at the dawn of technology}.
\newblock \bibinfo{journal}{Yearbook of Physical Anthropology}
  \bibinfo{volume}{47}, \bibinfo{pages}{118--164}.
%Type = Article
\bibitem[{Segre(2000)}]{seg00}
\bibinfo{author}{Segre, D.}, \bibinfo{year}{2000}.
\newblock \bibinfo{title}{Compositional genomes: Prebiotic information transfer
  in mutually catalytic noncovalent assemblies}.
\newblock \bibinfo{journal}{Proceedings of the National Academy of Sciences
  USA} \bibinfo{volume}{97}, \bibinfo{pages}{4112--4117}.
%Type = Article
\bibitem[{Sowden et~al.(2015)Sowden, Pringle and Gabora}]{sow}
\bibinfo{author}{Sowden, P.}, \bibinfo{author}{Pringle, A.},
  \bibinfo{author}{Gabora, L.}, \bibinfo{year}{2015}.
\newblock \bibinfo{title}{The shifting sands of creative thinking: Connections
  to dual process theory}.
\newblock \bibinfo{journal}{Thinking \& Reasoning} \bibinfo{volume}{21},
  \bibinfo{pages}{40--60}.
%Type = Article
\bibitem[{Tenenbaum et~al.(2011)Tenenbaum, Kemp, Griffiths and Goodman}]{tenen}
\bibinfo{author}{Tenenbaum, J.B.}, \bibinfo{author}{Kemp, C.},
  \bibinfo{author}{Griffiths, T.L.}, \bibinfo{author}{Goodman, N.D.},
  \bibinfo{year}{2011}.
\newblock \bibinfo{title}{How to grow a mind: statistics, structure and
  abstraction}.
\newblock \bibinfo{journal}{Science} \bibinfo{volume}{331 (6022)},
  \bibinfo{pages}{1279--1285}.
%Type = Incollection
\bibitem[{Veloz et~al.(2011)Veloz, Gabora, Eyjolfson and Aerts}]{vel}
\bibinfo{author}{Veloz, T.}, \bibinfo{author}{Gabora, L.},
  \bibinfo{author}{Eyjolfson, M.}, \bibinfo{author}{Aerts, D.},
  \bibinfo{year}{2011}.
\newblock \bibinfo{title}{Toward a formal model of the shifting relationship
  between concepts and contexts in different modes of thought}, in:
  \bibinfo{editor}{Song, D.}, \bibinfo{editor}{Melucci, M.},
  \bibinfo{editor}{Frommholz, I.}, \bibinfo{editor}{Zhang, P.},
  \bibinfo{editor}{Wang, L.}, \bibinfo{editor}{Arafat, S.} (Eds.),
  \bibinfo{booktitle}{Lecture Notes in Computer Science 7052: Proceedings of
  the Fifth International Symposium on Quantum Interaction}.
  \bibinfo{publisher}{Springer}, \bibinfo{address}{Berlin}, pp.
  \bibinfo{pages}{25--34}.
%Type = Article
\bibitem[{Vetsigian et~al.(2006)Vetsigian, Woese and Goldenfeld}]{vet06}
\bibinfo{author}{Vetsigian, K.}, \bibinfo{author}{Woese, C.},
  \bibinfo{author}{Goldenfeld, N.}, \bibinfo{year}{2006}.
\newblock \bibinfo{title}{Collective evolution and the genetic code}.
\newblock \bibinfo{journal}{Proceedings of the National Academy of Sciences
  USA} \bibinfo{volume}{103}, \bibinfo{pages}{10696--10701}.
%Type = Article
\bibitem[{W\"{a}chtersh\"{a}user(1997)}]{wac97}
\bibinfo{author}{W\"{a}chtersh\"{a}user, G.}, \bibinfo{year}{1997}.
\newblock \bibinfo{title}{The origin of life and its methodological challenge}.
\newblock \bibinfo{journal}{Journal of Theoretical Biology}
  \bibinfo{volume}{187}, \bibinfo{pages}{483--494}.

\end{thebibliography}

\section{Appendix: Justification of Proposition~\ref{mainpro}}

Let  $\cR_t$ be the set of actual updating reactions that occur in $W_t$ up to time $t$ (as noted, we assume that $t$ is large enough so this set is nonempty), and let $r_1, r_2, \ldots, r_k$ be a list of the sequence of these reactions in the order they occur.  The reactant of each reaction $r_i$ in $\cR_t$ is either an element of $M_0$ (which is a subset of $F$) or another attended item that is the product of an earlier reaction $r_j$ (where $j<i$) in the sequence (this holds even in the presence of attention shifts). Moreover, each reaction in $\cR_t$ is catalysed  either by an stimulus (which is an element of $F$) or by the product of an
earlier reaction $r_j$ (where $j<i$)  in the sequence.  Thus, $\cR_t$ satisfies the required properties to be a CAF (constructively autocatalytic F-generated set) as defined in \cite{mos}, and so is
a RAF. 

For the second part of Proposition~\ref{mainpro}, if $\lambda$ is sufficiently small then the rate of cognitive updating reactions will be less  than the rate $\mu$ at which items in working memory become no longer present in this set. By standard (birth-death process) arguments, it follows that a sequence of consecutive cognitive updating reactions (as in Fig.~\ref{ccpfig}(ii)) will eventually die out, and as $\lambda$ declines further the frequency of such sequences of length greater than 1 tends to zero.  However, as, $\lambda$ grows beyond the rate at which cognitive updating reactions exceeds  $\mu$,  sequences of consecutive cognitive updating reactions become increasingly frequent, and of greater duration and size.
\end{document}